\documentclass[10pt,letterpaper]{article}
\usepackage{opex3}

\usepackage{amsmath,amssymb}
\usepackage[utf8]{inputenc}
\usepackage[T1]{fontenc}%
\usepackage{dcolumn}%
\usepackage{bm}%
\usepackage{subfigure}%
\usepackage{hyperref}
\usepackage{xcolor}
\usepackage{subfigure}
\usepackage{cite}

\newcommand{\cm}{cm\ensuremath{^{-1}}~}

\newcommand{\beq}{\begin{equation}\begin{aligned}}
\newcommand{\eeq}{\end{aligned}\end{equation}}

\newcommand{\half}{\ensuremath{\frac{1}{2}\,}~}

\newcommand{\imag}{\ensuremath{\textrm{i}}}
\newcommand{\degree}{\ensuremath{^\circ\,}}
\newcommand{\subsubreffig}[1]{({\sffamily #1})}

\definecolor{linkcol}{rgb}{0,0,0.4} 
\definecolor{citecol}{rgb}{0.5,0,0} 

\hypersetup{
bookmarksopen=true,
pdftitle="Fabry-Perot enhanced Faraday rotation in graphene",
pdfauthor="N. Ubrig et al.", 
pdfsubject="Fabry-Perot enhanced Faraday rotation in graphene", 
pdfstartview={FitH},    
pdfmenubar=true, 
pdfhighlight=/O, 
colorlinks=true, 
pdfpagemode=UseNone, 
pdfpagelayout=SinglePage, 
pdffitwindow=true, 
linkcolor=linkcol, 
citecolor=linkcol, 
urlcolor=blue
}
\begin{document}


\title{Fabry-Perot enhanced Faraday rotation in graphene}

\author{Nicolas Ubrig,$^{1}$ Iris Crassee,$^{1}$ Julien Levallois,$^{1}$ Ievgeniia O. Nedoliuk,$^{1}$ Felix Fromm,$^{2}$ Michl Kaiser,$^{3}$ Thomas Seyller,$^{2}$ and Alexey B. Kuzmenko$^{1,\ast}$}

\address{$^{1}$Département de Physique de la matière condensée (DPMC), University of Geneva, Switzerland\\
$^{2}$Institut für Physik - Technische Physik, Technische Universität Chemnitz, 09126 Chemnitz, Germany, EU\\
$^{2}$Lehrstuhl f\"{u}r Werkstoffwissenschaften, Universit\"{a}t Erlangen-N\"{u}rnberg, 91058 Erlangen, Germany, EU}

\email{$^{\ast}$Corresponding author: alexey.kuzmenko@unige.ch}


\begin{abstract}
We demonstrate that giant Faraday rotation in graphene in the terahertz range due to the cyclotron resonance is further increased by constructive Fabry-Perot interference in the supporting substrate. Simultaneously, an enhanced total transmission is achieved, making this effect doubly advantageous for graphene-based magneto-optical applications. As an example, we present far-infrared spectra of epitaxial multilayer graphene grown on the C-face of 6H-SiC, where the interference fringes are spectrally resolved and a Faraday rotation up to 0.15 radians (9$\degree$) is attained. Further, we discuss and compare other ways to increase the Faraday rotation using the principle of an optical cavity.
\end{abstract}




\section{Introduction}

Graphitic materials, such as carbon nanotubes and graphene, find numerous applications in various fields of optics \cite{ren_carbon_2009,yan_dual-gated_2012,gabor_hot_2011,xia_ultrafast_2009}. The giant terahertz Faraday rotation in graphene \cite{crassee_giant_2011,ferreira_faraday_2011,fialkovsky_faraday_2012,da_enhanced_2011,fallahi_manipulation_2012,shimano_quantum_2013} suggests that this novel material can be useful in  applied magneto-optics. The exceptionally strong Faraday effect, combined with a high doping tunability, which is a hallmark of graphene, may potentially lead to a new class of ultrafast tunable magneto-optical modulators and isolators \cite{crassee_giant_2011,da_enhanced_2011,da_graphene-based_2012,zhou_tunable_2013}. Apart from practical importance, the magneto-optical phenomena are helpful to study the charge carrier dynamics in these systems\cite{crassee_giant_2011,crassee_multicomponent_2011,levallois_decrypting_2012}.

Even though the observed Faraday angles of a few degrees at fields of only a few Tesla \cite{crassee_giant_2011} are exceptionally large for a single atomic layer, the use of this effect in practical devices will certainly be facilitated by increasing the rotations even more, while reducing the required magnetic field. Modifying the properties of graphene itself, for example, increasing the number of magneto-optically active layers, enhancing the mobility of charge carriers and fabrication of plasmonic nanostructures, is obviously one of the avenues for this improvement. However, the electromagnetic properties of the environment surrounding a magneto-optical layer, also play an important role. In particular, it is known that small Faraday rotation can be boosted by placing magneto-optically active samples inside a Fabry-Perot cavity. The rotation angle is enhanced after multiple internal beam passages \cite{rosenberg_resonant_1964}. This principle was used to build magneto-optical devices \cite{stone_
enhancement_1990}, to increase the sensitivity of magnetic field sensors \cite{wagreich_magnetic_1996} or to measure ultrasmall Verdet constants \cite{jacob_small_1995}.  In the context of graphene, the idea was recently studied theoretically in \cite{ferreira_faraday_2011}, Ferreira et al.

An important issue to be addressed in this context is the influence of a substrate, which is always present in realistic applications. One undesirable effect of the substrate is to reduce the magneto-optical rotation as compared to free standing graphene by a factor, which depends on the substrate refractive index \cite{crassee_giant_2011}. However, a well polished flat parallel substrate can also develop strong Fabry-Perot interference and thus acts as a cavity. In this work we show that it can be used to increase the rotation angle and simultaneously the total transmission. In order to make a connection between this effect and previous ideas about the Fabry-Perot cavity, we perform several model calculations. This allows us to compare different ways to enhance the Faraday rotation in graphene.

\section{Experimental Details}

\begin{figure}
 \centering
 \includegraphics[width=0.45\linewidth]{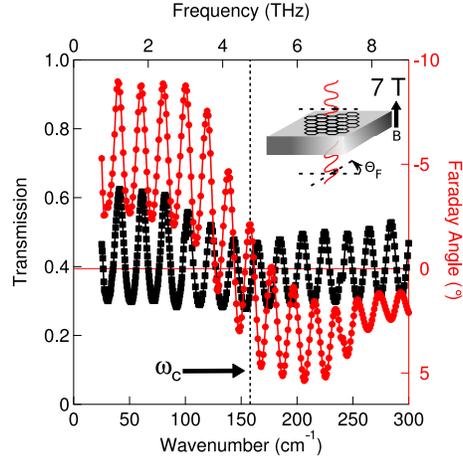}
 \caption{Transmission (black squares, left axis) and Faraday rotation (red circles, right axis) spectra at 7 Tesla and 5 Kelvin.
 Note that the Faraday rotation scale is inverted. The horizontal line corresponds to zero Faraday rotation, the vertical line indicates the cyclotron frequency.}
 \label{fig_TRaFR}
\end{figure}

Our sample is multilayer epitaxial graphene grown on the C-face\cite{berger_ultrathin_2004} of 6H-SiC by annealing silicon carbide in an Ar atmosphere for 90 minutes at 1650 $\degree$C. Before the growth, the substrate was hydrogen-etched at 1600$\degree$C for 15 minutes. The total number of graphene layers was about 20, as determined by X-ray photoelectron spectroscopy  and confirmed by infrared absorption in the near-infrared range. However, only one or two layers closest to the SiC are highly doped (n-type) \cite{berger_electronic_2006} and therefore responsible for the classical cyclotron resonance, which dominates the magneto-optical response in the range of energies and magnetic fields \cite{crassee_multicomponent_2011} considered in this paper. The remaining layers, which are quasineutral, only contribute to the overall absorption at low frequencies \cite{sadowski_landau_2006,crassee_multicomponent_2011}. The thickness of the substrate was reduced to $d$ = 80 $\mu$m by polishing in order
to increase the period of the Fabry-Perot oscillations (about 20 \cm in the THz range). We measured magneto-transmission $T(\omega, B)$ with unpolarized light and the Faraday rotation $\theta_F(\omega, B)$ by a two-polarizer technique between 15 and 700 \cm with the help of a Fourier transform spectrometer coupled to a split-coil superconducting magnet \cite{crassee_giant_2011}. A mercury light source, silicon beamsplitter and liquid-helium cooled Si bolometer were used. The spectral resolution (1 \cm for the transmission and 2 \cm for the Faraday rotation) was sufficient to fully resolve the interference fringes. The spectra at 5 Kelvin and 7 Tesla are plotted in Fig. \ref{fig_TRaFR}. At lower fields the data show a qualitatively similar behavior. The energy of the cyclotron resonance (marked by the dashed line) can be identified as a broad minimum in (oscillation-averaged) transmission and a reduced amplitude of the fringes. The (oscillation-averaged) Faraday angle changes sign in this 
region. The
negative value at low frequencies indicates that the layer responsible for the resonance is n-doped.

\section{Discussion}

An important observation is that the maxima of the transmission and the absolute value of the Faraday rotation virtually coincide, except close to the cyclotron resonance, $\omega_c$. It means that a constructive interference between internally reflected beams is favorable for both quantities. We note also that at some frequencies the Faraday angle reaches 9$\degree$, which is 50 percent higher than the value reported earlier\cite{crassee_giant_2011}.

In order to explain this finding, we model the experimental spectra by treating all graphene layers as a thin film with the total conductivity $\sigma_{+}$ ($\sigma_{-}$) for the right- (left-) circular polarized light:

\beq
\sigma_{\pm}(\omega)  = \frac{2D}{\pi}  \frac{i}{\omega \mp\omega_c+i\gamma} + \sigma_b
\label{EqSpm2}
\eeq

\noindent represented by a sum of the cyclotron resonance, described by a Drude weight $D$, a cyclotron frequency $\omega_c$ and a scattering rate $\gamma$, and a constant background $\sigma_b$, approximating the absorption in other layers\cite{crassee_multicomponent_2011,CrasseeFuture}. This quasi-classical approach is valid in our case since the condition $\omega,\omega_c < 2E_F$ is satisfied\cite{gusynin_universal_2009,ferreira_faraday_2011}. The transmission and the Faraday rotation are given by:
\beq
T = \frac{|t_{-}|^2 + |t_{+}|^2}{2},\quad \theta_F = \half\mathrm{arg}\left(\frac{t_{-}}{t_{+}}\right)
\label{eq:TF}
\eeq
\noindent where $t_{\pm}$ are the complex transmission coefficients for the two circular polarizations. Taking into account multiple internal reflections in the substrate, for which the refractive index $n \approx 3.1$ and zero absorption were assumed in the spectral range of interest we get\cite{heavens_thin_1970}:

\beq
t_{\pm} = 4n \tau_s \cdot \Big[ (n+1)^2 - (n-1)^2\tau_s^2 + Z_0\sigma_{\pm} \left( n + 1 + (n-1)\tau_s^2 \right) \Big]^{-1},
\label{eq:complext}
\eeq

\noindent where $\tau_s = \exp(\imag \omega d n/c)$ is the phase factor acquired by light in the substrate and $Z_0$ is the impedance of vacuum.

\begin{figure}
 \centering
 \includegraphics[width=0.37\linewidth]{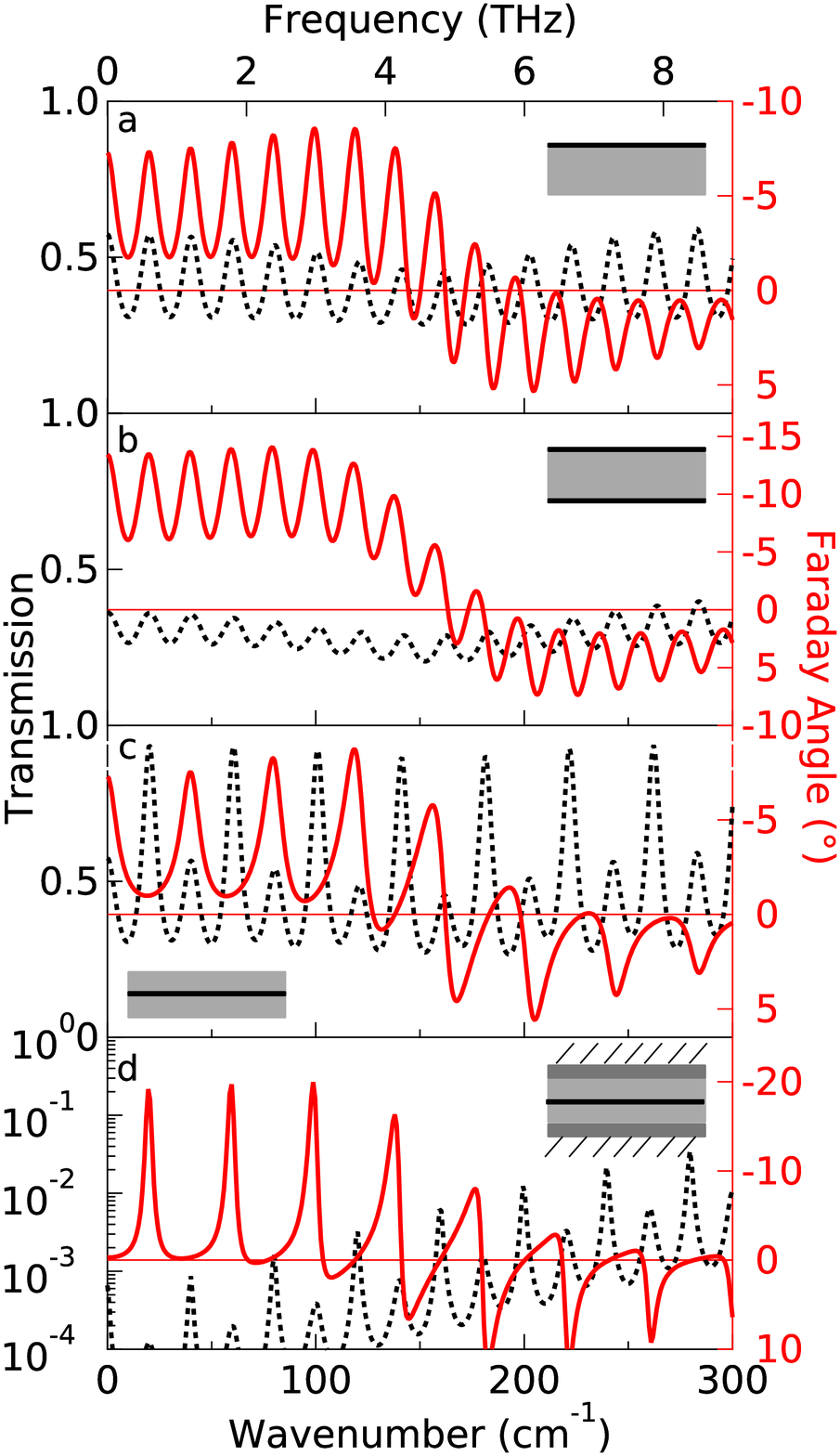}
 \caption{Simulation of the magneto-optical transmission and Faraday rotation of graphene on and in SiC for three different configurations. Note that the Faraday rotation scale is inverted. \subsubreffig{a} Graphene covers one side of SiC, \subsubreffig{b} graphene covers both sides of SiC, \subsubreffig{c} graphene is in the middle of SiC slab, \subsubreffig{d} graphene is in the middle of a SiC slab covered by metallic layers on both sides.}
 \label{fig_Xpos}
\end{figure}

The following parameter values were found to fit the data at 7 Tesla in the best way: $\omega_c =$ 156 \cm , $\gamma =  $56 \cm, $D/\sigma_0$ = 3620 \cm, $\sigma_b/\sigma_0 = 23$, where $\sigma_0=e^2/4\hbar$ is the universal conductivity of monolayer graphene \cite{ando_dynamical_2002}. We estimate the Fermi energy $E_F$ = 0.24 eV from the Drude weight extracted from the fit through the relation $D = 2\sigma_0 E_F / \hbar$. The theoretical curves are shown in Fig. \ref{fig_Xpos}\subsubreffig{a}. One can see that this simple model reproduces the data accurately, including the coincidence of maxima of the Faraday rotation and transmission.

The physical meaning of this result  becomes obvious in the following. The constructive interference occurs at frequencies, where $\tau_s = \pm 1$. In this case, Eq. (\ref{eq:complext}) reduces to

\beq
t_{\mathrm{constr}, \pm} = \tau_s\left(1 + \frac{Z_0\sigma_{\pm}}{2}\right)^{-1},
\label{topt}
\eeq

\noindent which is equal, apart from the prefactor $\tau_s$, to the transmission of free standing graphene with the same optical conductivity\cite{nair_fine_2008,kuzmenko_universal_2008}. It thus appears that the effect of the constructive Fabry-Perot interference is to exactly compensate the screening effect of the substrate. If monochromatic light is used (such as a terahertz laser), it would be desirable in magneto-optical applications to adjust the substrate thickness in order to achieve this condition.

Developing on this result, we next explore theoretically other possibilities to improve the Faraday rotation in graphene by making use of the cavity principle. For the sake of simplicity, we assume that graphene has the same optical conductivity as the one obtained in the present experiment and the total thickness of the substrate is always the same. In this case the substrate can be formally regarded as a low-finesse ($F\approx 3$) optical cavity created by the SiC-vacuum interfaces.

We first consider the case where both sides of the substrate are covered with graphene. Although this cannot be done by growing graphene on two sides one can simply press two identical samples together. Equation (\ref{eq:complext}) is now modified as follows:

\beq
t_{\pm} = 4n \tau_s \cdot \Big[ (n + 1)^2-(n - 1)^2\tau_s^2 + 2Z_0\sigma_{\pm} \left( n + 1 + (n-1)\tau_s \right)^2 + Z_0^2\sigma_{\pm}^2(1 - \tau_s^2) \Big]^{-1}.
\label{eq:complextoutsides}
\eeq

\noindent For the constructive interference we obtain $t_{\mathrm{constr}, \pm} = \tau_s\left(1 + Z_0\sigma_{\pm}\right)^{-1}$, which is the same result as Eq. (\ref{topt}), except for the factor 2 in the graphene conductivity term, due to the presence of two graphene layers. The computed spectra are plotted in Fig. \ref{fig_Xpos}\subsubreffig{b}. The Faraday rotation now reaches 15\degree. The transmission is lowered but remains at a reasonable level.

Next, we consider the graphene film to be in the middle of a SiC plate, which can, in principle, be achieved by pressing a substrate with graphene against a bare substrate with the same thickness. This case is described by the equation:

\beq
t_{\pm} = 4n \tau_s \cdot \Big[ (n + 1)^2-(n - 1)^2\tau_s^2 + \frac{Z_0\sigma_{\pm}}{2n}\left(n + 1+(n - 1)\tau_s\right)^2 \Big]^{-1} \, .
\label{eq:complext2}
\eeq

\noindent The computed spectra are plotted in Fig. \ref{fig_Xpos}\subsubreffig{c}. Now the cases $\tau_s = +1$ and -1 are fundamentally different. In the first case the result for free-standing graphene (Eq. (\ref{topt})) is again recovered. Indeed, the same transmission and rotation as in Fig. \ref{fig_Xpos}\subsubreffig{a} are observed for these frequencies. However, if $\tau_s = -1$ then $t_{\pm} = \tau_s(1 + Z_0\sigma_{\pm}/2n^2)^{-1}$, which means that the effective conductivity of graphene is reduced by a factor of $n^2\approx 10$ as compared to Eq. (\ref{topt}). Although this case is beneficial for the overall transmission, the Faraday rotation is so small that this case is not of practical interest for magneto-optical applications. Thus, the configuration with graphene in the middle is not more advantageous for the enhancement of the Faraday rotation than the original one.

Finally, the SiC slab in the previous configuration can be turned into a high-finesse Fabry-Perot cavity by coating each of the SiC-vacuum interfaces with a highly reflecting metallic layer. This can be modeled by substituting $n\pm 1\rightarrow n\pm 1\pm Z_0\sigma_m$ in Eq. (\ref{eq:complext2}), where $\sigma_m$ is the sheet optical conductivity of the deposed metal, which we assume to be magneto-optically inactive. In Fig. \ref{fig_Xpos}\subsubreffig{d} we present such a calculation for a metallic layer with a Drude conductivity of $\sigma_m(\omega) = \sigma_{\mathrm{DC}}/(1-i\omega\tau)$, a static value $\sigma_{\mathrm{DC}}$ = 0.1 $\Omega^{-1}$ and a scattering rate $1/\tau = 100$ \cm. The finesse of the cavity is about 10 in the considered spectral range. Now the Faraday rotation reaches about 20 degrees at some frequencies. This is more than twice higher than the maximum achievable value in the uncoated cavity (Fig. \ref{fig_Xpos}\subsubreffig{c}), which is a 
manifestation of the cavity boost \cite{ferreira_faraday_2011}. Although the cavity principle works, the obvious penalty is that the total transmission is significantly reduced because of the low transmission of the metallic mirrors.
By comparing Figs. \ref{fig_Xpos}\subsubreffig{b} and \subsubreffig{d} we conclude that placing graphene on both sides of a low-finesse cavity, {\em i.e.} increasing the number of magneto-optically active layers, is a better strategy to increase the Faraday rotation without drastically reducing the transmission.

One should note that the magneto-optically inactive graphene layers in the present sample, modeled with the term $\sigma_b$, only reduce the transmission without improving the rotation. In monolayer graphene grown on the Si-face \cite{emtsev_towards_2009,riedl_quasi-free-standing_2009}, such layers are absent and the transmission is expected to be higher than in the data presented here, with a similar value of the Faraday rotation.

In conclusion, we report a significant enhancement of the Faraday rotation in graphene due to constructive Fabry-Perot interference in the substrate as compared to the case where the interference is not resolved. We show that under these conditions the total transmission of the system is also increased. This observation, as well as our simulations of the Faraday rotation and transmission in related graphene-cavity systems, contribute to a possible use of graphene for novel magneto-plasmonic applications. Even though the absolute rotation angles achievable with graphene do not compete at the moment with the values obtained with the aid of conventional thick magneto-optical materials, the potential possibility to tune the effect by electrostatic gating opens avenues for new functionalities, such as an ultrafast inversion of the rotation angle by electric field. The results obtained in this work may also be of interest when combined with a lasing medium inside the cavity \cite{morimoto_cyclotron_2009}.

\section*{Acknowledgments}
We acknowledge Daniel Chablaix, Michaël Tran and Mehdi Brandt for technical assistance. This work was supported by the SNSF through project 140710 and via the NCCR "Materials with Novel Electronic Properties - MaNEP". We acknowledge support by the EC under Graphene Flagship (contract no. CNECT-ICT-604391).

\end{document}